\documentclass[fleqn,nobibnotes]{revtex4}

\usepackage{amsmath}
\usepackage{graphicx}

\def\lsim{\raise0.3ex\hbox{$<$\kern-0.75em\raise-1.1ex\hbox{$\sim$}}}
\def\gsim{\raise0.3ex\hbox{$>$\kern-0.75em\raise-1.1ex\hbox{$\sim$}}}

\begin{document}

\title{What kind of energy is the mass?}
\author{E. R. Cazaroto}
\email{ecazaroto@yahoo.com.br}
\affiliation{
Instituto de F\'{\i}sica, Universidade de S\~{a}o Paulo,
C.P. 66318,  05315-970 S\~{a}o Paulo, SP, Brazil\\
}

\begin{abstract}
In 1905, Einstein discovered the famous equation: $E=mc^2$, which means that the rest mass of a particle is some kind of energy. This energy is generally referred to as ``rest energy'', since the particle is believed to be at rest. This paper proposes a new interpretation to the term $mc^2$. Observing the similarity between the term $mc^2$ and the kinetic energy term $mv^2/2$, we propose to interpret $mc^2$ as being one term of kinetic energy. In other words we propose that, in the called ``rest frame'', the massive particles are not really at rest, but they are doing a special kind of motion at the light speed c. In this interpretation the ``mass'' is not an intrinsic property of the particle. The ``mass'' is simply the kinetic energy associated with this special kind of motion. The more important consequence of this hypothesis is that the term $mc^2$, present in the relativistic Hamiltonian, must be rewritten as: $mc^2 \rightarrow p_0 \, c$, where $p_0 = mc$ is the modulus of the instantaneous linear moment $\vec{p}_0$ associated with the special motion. We analyse a physical scenario in which this special motion is a Microscopic Orbital Circular Motion (MOCM). The characteristic length scale of the MOCM is of the order of the Compton wavelength of the particle.
\end{abstract}

\maketitle

\section{Introduction}

Since the discovery of the formula $E=mc^2$ by Einstein, we know that the inertial mass of a particle at rest corresponds to the total amount of energy stored in the particle. It includes all forms of energy. For example, if the particle is not fundamental, being constituted by other particles, its mass is not the simple sum of the individual masses of its constituents, but the energy of interaction between them is also included in the sum, contributing to the total mass of the particle. There are, however, massive particles in nature, for example the electron, that are believed to be fundamental, i.e., massive particles that do not have internal structure. The question that we make here is: what kind of energy is the mass of a fundamental particle?

The recent discovery of the Higgs Boson in LHC is not the final point in what concerns the mass origin. Although the mass generation of the gauge bosons that intermediate the weak interaction is naturally described by the Higgs mechanism, the fermion masses are ``put on hand'' in the Lagrangian. Therefore, we are not sure if the Higgs mechanism is the one that give mass to the charged leptons. Furthermore, even if the Higgs mechanism is responsible for generating the lepton masses, there is not yet a convincing explanation to the fact that the Yukawa coupling between the Higgs boson and each charged lepton is so different. The charged leptons have the same quantum numbers, except by their masses which are different. This fact often leads theoretical physicists to think on the muon and on the tau as being excited states of the electron. However, there is not yet a satisfactory model to describe such excited states, and new ideas that could shed light and inspire the construction of a successful theory towards this direction are welcome. 

We propose in this paper a different point of view about the nature of the rest mass that could be a starting point towards the construction of a theory in which some fundamental massive particles could be thought as being excited states of other ones. We propose that the term $mc^2$ is a term of kinetic energy, which is associated with a special kind of motion, done by all the massive fundamental particles, at the light speed $c$. 

The paper is organized as follows. In Section \ref{rest} we propose and justify the interpretation of $mc^2$ as being a term of kinetic energy. In Section \ref{outrest} we extend this interpretation to the case where the particle is observed from other inertial frames. In Section \ref{semi_classical} we analyse a physical scenario in which the special microscopic motion, whose kinetic energy is $mc^2$, is a Microscopic Orbital Circular Motion (MOCM). Finally, in Section \ref{conc} we make the summary and the final conclusions.

\section{$mc^2$: one term of kinetic energy}
\label{rest}

The term $mc^2$ has the same structure as the term of kinetic energy $mv^2/2$. What we mean is that these two terms have the structure ``mass times the square of the velocity''. It means that, if $mv^2/2$ represents the kinetic energy of a particle with mass $m$ and velocity $v$, then the term $mc^2$ can be interpreted as being not the rest energy, but the kinetic energy of a particle with mass $m$ and velocity $c$.

It is possible to argue that ``{\it the light speed $c$ in the vacuum is a mere constant that gives the dimension of energy to the mass in the MKS system}''. This argument, however, is not true, i.e., the light speed $c$ is not a mere constant. In first place, the light speed only has the same value $c$ in any inertial frame because the time and the space are relative quantities, that depend on the referential. It can be considered a mere coincidence the fact that the Lorentz transformations, for the time and for the space, have exactly the required shape in order that the light speed in the vacuum remains with the same value $c$ in any inertial frame. Second, although $c$ has the same value in any inertial frame, we must remember that $c$ is a function of $\epsilon_0$ and $\mu_0$, respectively the electric and magnetic permeabilities of the vacuum, through the formula \cite{eletro}:
\begin{equation}
c = \frac{1}{\sqrt{\epsilon_0\mu_0}} \, .
\end{equation}
So the light speed $c$ in the vacuum has the particular value $c \approx 3\times10^8 m/s$ because the permeabilities $\epsilon_0$ and $\mu_0$ have the particular values that they have. However, the values of $\epsilon_0$ and $\mu_0$ could be different than they are, what would imply in a different value for $c$. In fact, in a material medium, where the permeabilities $\epsilon$ and $\mu$ are different of $\epsilon_0$ and $\mu_0$, the electromagnetic waves propagate with velocity different of $c$ \cite{eletro}.

Other possible question is: ``{\it why can we compare the term $mc^2$, product of relativity theory, with the non-relativistic term $mv^2/2$ ?}''. The answer for this question is that although the term $mc^2$ was discovered in special relativity theory, this term is the first one to survive in the non-relativistic limit, i.e., in the limit $v<<c$. In fact, when we make the expansion of the relativistic energy, the zero order term of the expansion is the term $mc^2$. In this expansion, the term $mv^2/2$ arises as a first order correction for the energy:
\begin{equation}
E = \frac{mc^2}{\sqrt{1-v^2/c^2}} \approx mc^2 + \frac{mv^2}{2} + ... \, .
\end{equation}
In this sense, we can say that the term $mc^2$ is more ``non-relativistic'' than the term $mv^2/2$. In short, we can say that these two terms are important in the same kinematical regime, i.e., in the regime $v<<c$, and they can be compared between themselves. 

In summary, we are proposing to interpret the term $mc^2$ as being a term of kinetic energy, which is associated with a special kind of motion done by all the massive fundamental particles at the velocity of light. Obviously, this special motion cannot be a macroscopic motion in the 3-dimensional space, otherwise we would see the massive particles travelling through the space like the massless ones. The fact that we see the massive particles ``at rest'' indicates that this special motion happens in a microscopic scale. 

Finally, the fact that the massive fundamental particles move at the light speed $c$ implies that the physical property called ``mass'' is not an intrinsic property of the particles. According to special relativity theory, only massless particles can move at the light speed $c$. Therefore, all the fundamental particles that exist in nature are intrinsically massless. In the present theory, the only difference between a particle classified as ``massive'' and a particle classified as ``massless'' is the type of motion that the particle does. If the particle moves rectilinearly through the space, at the light speed $c$, it is classified as ``massless''. On the other hand, if the particle makes the special microscopic motion proposed here, also at the light speed $c$, it is classified as ``massive''. In Section \ref{semi_classical} we will analyse a physical scenario in which this special motion is a Microscopic Orbital Circular Motion.

\subsection{The linear moment associated with $mc^2$}
\label{pmc}

The energy $E$ of a massless particle is related with its linear moment $p$ through the expression:
\begin{equation}
E = p \, c \, .
\label{disp_rel}
\end{equation}
Since in the present theory the ``mass'' is not an intrinsic property of the particle, the relation (\ref{disp_rel}) is valid also for ``massive'' particles. Let us label the linear moment of a ``massive'' particle as $\vec{p}_0$. Since the particle is doing a special microscopic motion, $\vec{p}_0$ is the instantaneous linear moment associated with this motion. Using relation (\ref{disp_rel}), the energy associated with this special motion can be rewritten as a function of $p_0$:
\begin{equation}
E = mc^2 = p_0 \, c \, ,
\label{disp_rel_mass}
\end{equation}
where $p_0$ is the modulus of the vector $\vec{p}_0$. Therefore, looking at Eq. (\ref{disp_rel_mass}) we conclude that:
\begin{equation}
p_0 = mc \, .
\label{p0_mc}
\end{equation}

\subsection{Comparing $mc^2$ with $mv^2/2$}

Now let us understand better the differences between $mc^2$ and $mv^2/2$ when we interpret $mc^2$ as being one term of kinetic energy. 

First, observe that, in general, when we take the kinetic energy $E = mv^2/2$, the mass ($m$) of the particle is kept fixed, and each value $E$ of the energy corresponds to a different value $v$ of the velocity. In particular, if the system is quantized (suppose, for example, that the particle is inside a potential well) the discrete energies $E_i$ are related with the discrete velocities $v_i$ through the expression: 
\begin{equation}
E_i = \frac{m v_i^2}{2} \,\,\,\,\,\, , \,\,\,\,\,\, i = 1,2,3,... \,\,\, .
\end{equation}
On the other hand, the light speed $c$ has a fixed value. Therefore, when we take the expression $E = mc^2$ as being the kinetic energy of the particle, each different energy $E$ must corresponds to a different mass $m$. In particular, if the system is quantized, the discrete energies $E_i$ correspond to the discrete masses $m_i$ through the formula: 
\begin{equation}
E_i = m_ic^2 \,\,\,\,\,\, , \,\,\,\,\,\, i=1,2,3,... \,\, . 
\label{eimic2}
\end{equation}

Now let us calculate the energy acquired by a particle that is under the action of a force $\vec{F}$. When the force is applied on the particle, its energy $E$ will change an amount $\Delta E$ given by:
\begin{equation}
\Delta E = \int \vec{F} \cdot d\vec{r} = \int \frac{d\vec{p}}{dt}\cdot d\vec{r} = \int d\vec{p}\cdot \frac{d\vec{r}}{dt} = \int d\vec{p} \cdot \vec{v} \,\,\, .
\label{enerchang}
\end{equation}
In a mathematically more rigorous way we can do the last two passages as:
\begin{equation}
\frac{d\vec{p}}{dt}\cdot d\vec{r} = lim_{\Delta t \rightarrow 0} \,\,\, \left\{ \frac{\Delta \vec{p}}{\Delta t}\cdot \Delta \vec{r} \right\} = lim_{\Delta t \rightarrow 0} \,\,\, \left\{ \Delta \vec{p} \cdot \frac{\Delta \vec{r}}{\Delta t} \right\} = d\vec{p} \cdot \vec{v} \,\,\, .
\end{equation}
The particle with kinetic energy $E = mv^2/2$ has a moment $\vec{p} = m\vec{v}$. An infinitesimal change in $\vec{p}$ is given by: $d\vec{p} = md\vec{v}$. Substituting $d\vec{p} = md\vec{v}$ in Eq. (\ref{enerchang}), and considering that initially the particle is at rest, we obtain that the energy $E$ acquired by the particle is given by:
\begin{equation}
E = \int d\vec{p} \cdot \vec{v} = \int (m d\vec{v}) \cdot \vec{v} = m \int \vec{v}.d\vec{v} = \frac{mv^2}{2} \,\,\, .
\label{deduzmv2s2}
\end{equation}
Therefore, by substituting the infinitesimal change $d\vec{p} = md\vec{v}$ in Eq. (\ref{enerchang}), we were able to obtain the expression of the kinetic energy $E = mv^2/2$.

Now let us consider the expression $E = mc^2$ for the kinetic energy. As discussed in the previews subsection, the linear moment associated with this energy is: $\vec{p}_0 = m\vec{c}$. But now, the modulus of the velocity $\vec{c}$ has a fixed value $c$. It means that the vector $\vec{c}$ determines only the direction of the vector $\vec{p}_0$. The magnitude of $\left| \vec{p}_0 \right|$ is determined by the ``mass'' $m$, which in this case is not constant. By applying a force on the particle, we cannot change the magnitude of the vector $\vec{c}$, but only the direction of this vector. Therefore, if we apply a force $\vec{F}$ parallel to $\vec{c}$, we will have $\Delta \vec{c} = \vec{0}$ and $\Delta m \neq 0$. In this case, the change in the linear moment of the particle is given by: $\Delta \vec{p}_0 = (\Delta m)\vec{c}$. On the other hand, if the force is applied in any other direction we have: $d\vec{p}_0 = (dm)\vec{c} + m(d\vec{c})$. But since $\left| \vec{c} \right| = $ constant, the variation $d\vec{c}$ can happens only perpendicularly to the velocity $\vec{c}$, what means that $\vec{c} \cdot (d\vec{c}) = 0$. 

Substituting $d\vec{p}_0 = (dm)\vec{c} + m(d\vec{c})$ in Eq. (\ref{enerchang}), and considering that $\vec{c} \cdot (d\vec{c}) = 0$, we have:
\begin{equation}
E \,\, = \,\, \int d\vec{p}_0 \cdot \vec{v} \,\, = \,\, \int (dm \, \vec{c} + m(d\vec{c})) \cdot \vec{c} \,\, = \,\, \int dm (\vec{c} \cdot \vec{c}) \,\, = \,\, c^2 \int dm \,\, = \,\, mc^2 \,\,\, .
\label{deduzmc2}
\end{equation}
Therefore, doing the same procedure as in the previews case, we were able to obtain the expression of the kinetic energy $E=mc^2$.

Comparing (\ref{deduzmc2}) with (\ref{deduzmv2s2}) it becomes clear why the term $mc^2$ does not have a factor $2$ in the denominator, like the term $mv^2/2$. This factor arises in the denominator when we make the integral of $\,\,\, \vec{v} \cdot d\vec{v} \,\,\,$ in (\ref{deduzmv2s2}), whereas in (\ref{deduzmc2}) we have the integral of $\, dm \,$.

\section{The massive particle out of its ``rest'' frame}
\label{outrest}

The formula $E = mc^2$ is valid only when the particle is observed in its ``rest frame''. According to special relativity theory, when the particle is moving with uniform rectilinear velocity $\vec{v}$ its energy is given by \cite{relativ}:
\begin{equation}
E = \frac{mc^2}{\sqrt{1-v^2/c^2}}  \,\,\, .
\label{ener3}
\end{equation}
Note that the velocity $\vec{v}$ appearing in this formula has nothing to do with the velocity of the special motion proposed in Section \ref{rest}. The velocity $\vec{v}$ appearing in Eq. (\ref{ener3}) is the macroscopic velocity of the particle. Observe that the special relativity theory was developed considering only the existence of the macroscopic motion, so that the velocity of the special microscopic motion is not explicit in the formulas of special relativity theory, for example it is not explicit in formula (\ref{ener3}).

The relativistic energy (\ref{ener3}) can be rewritten as a function of the macroscopic moment $\vec{p}$ of the particle. According to special relativity theory, $\vec{p}$ is given by:
\begin{equation}
\vec{p} = \frac{m\vec{v}}{\sqrt{1-v^2/c^2}} \,\,\, .
\label{momen1}
\end{equation}
Using this relation, the expression (\ref{ener3}) can be rewritten as:
\begin{equation}
E = \sqrt{(mc)^2 + p^2} \,\,  c  \,\,\, ,
\label{ener5}
\end{equation}
where $p$ is the modulus of $\vec{p}$. Since this expression does not depend explicitly on the velocity $\vec{v}$, in special relativity theory it is assumed to be the Hamiltonian $H$ of the free particle, i.e.:
\begin{equation}
H = \sqrt{(mc)^2 + p^2} \,\,  c  \,\,\, .
\label{ener7}
\end{equation}

In what follows we are going to identify, in the Hamiltonian (\ref{ener7}), the term that corresponds to the microscopic moment of the massive particle, namely the moment that is associated with the special microscopic motion proposed in previous section.

\subsection{Rewriting the relativistic Hamiltonian}

Making the macroscopic moment $\vec{p} = \vec{0}$ in Eq. (\ref{ener5}) we obtain:
\begin{equation}
E = (mc) \,  c  \,\,\, .
\label{ener9}
\end{equation}
As discussed in Subsection \ref{pmc}, the term `` $mc$ '' in Eq. (\ref{ener9}) is interpreted as the modulus of the microscopic moment $\vec{p}_0$ [see Eq. (\ref{disp_rel_mass})]. Therefore, we identify the term $\, mc \,$ in Eq. (\ref{ener5}) as being $\, p_0 \,$, and the relativistic energy (\ref{ener5}) is rewritten as:
\begin{equation}
E = \sqrt{p_0^2+ p^2} \,\, c  \,\,\, .
\label{ener11}
\end{equation}
Consequently, the relativistic Hamiltonian (\ref{ener7}) becomes:
\begin{equation}
H = \sqrt{p_0^2 + p^2} \,\,  c  \,\,\, .
\label{ener13}
\end{equation}

\subsection{The total moment of a ``massive'' fundamental particle}
\label{totmomen}

The structure of the Hamiltonian (\ref{ener13}) indicates that the microscopic moment $\vec{p}_0$ is perpendicular to the macroscopic moment $\vec{p}$. In fact, we can define the total moment $\vec{p}_{tot}$ of a ``massive'' fundamental particle as the sum of the macroscopic moment $\vec{p}$, given by Eq. (\ref{momen1}), with the microscopic moment $\vec{p}_0$:
\begin{equation}
\vec{p} _{tot} = \vec{p}_0 + \vec{p} \,\,\, ,
\label{ptot}
\end{equation}
where $|\vec{p}_0| = mc $. 

As discussed in subsection \ref{pmc}, the fact that in the present theory the ``mass'' is not an intrinsic characteristic of the particle implies that the relation (\ref{disp_rel}), valid for massless particles, is valid also for ``massive'' particles. Therefore, in the present theory we have that the energy $E$ of a ``massive'' particle is related with its total moment $p_{tot}$ through:
\begin{equation}
E = p_{tot} \, c \, .
\label{ener15}
\end{equation}
Now, comparing Eq. (\ref{ener11}) with the relation (\ref{ener15}), we identify the modulus of the total moment as:
\begin{equation}
p_{tot} = \sqrt{p_0^2+ p^2} \, .
\label{mptot}
\end{equation}
Squaring the vector $\vec{p}_{tot}$, given by Eq. (\ref{ptot}), we obtain:
\begin{equation}
(\vec{p}_{tot}) ^2 = (\vec{p}_0 + \vec{p})^2 = p_0^2 + p^2 + 2\vec{p}_0 \cdot \vec{p} \,\,\,\,\,\,\,\,\,\,\,\,\,\,\,\,\,\, \Longrightarrow \,\,\,\,\,\,\,\,\,\,\,\,\,\,\,\,\,\, p_{tot} = \sqrt{p_0^2 + p^2 + 2 \vec{p}_0 \cdot \vec{p}} \,\,\, ,
\end{equation}
where $p_{tot} = |\vec{p}_{tot}|$. Comparing this result with Eq. (\ref{mptot}) we see that: $\vec{p}_0 \cdot \vec{p} = 0$, i.e., $\vec{p}_0$ is always perpendicular to $\vec{p}$.

\subsection{The total velocity of the ``massive'' fundamental particle}

After defining the total moment $\vec{p}_{tot}$ of the ``massive'' fundamental particle, Eq. (\ref{ptot}), we define now the total velocity $\vec{v}_{tot}$. Using the same notation as for the moment, we define the total velocity as:
\begin{equation}
\vec{v}_{tot} = \vec{v}_0 + \vec{v} \, ,
\label{vtot}
\end{equation}
where $\vec{v}_0$ corresponds to the microscopic special motion and $\vec{v}$ is the usual macroscopic velocity. As discussed above, the microscopic moment $\vec{p}_0$ is perpendicular to the macroscopic moment $\vec{p}$. Consequently, we have that the microscopic velocity $\vec{v}_0$ is perpendicular to the macroscopic velocity $\vec{v}$. 

\begin{figure}[htbp]
\includegraphics[scale=0.50]{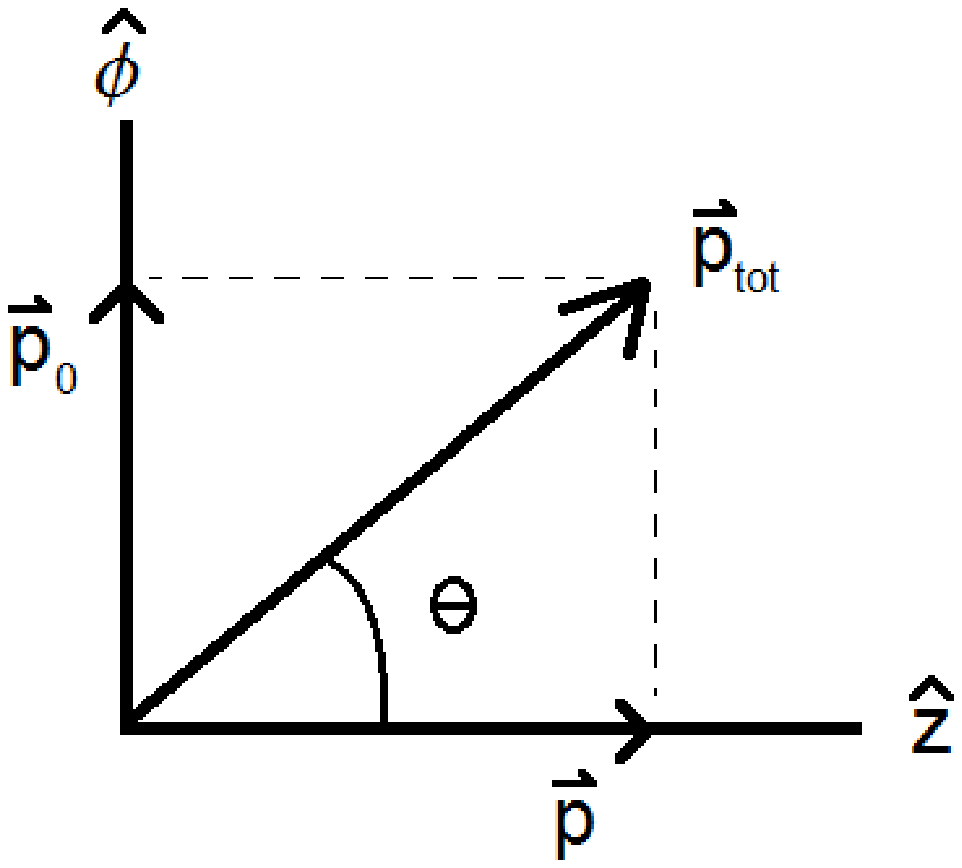}
\includegraphics[scale=0.50]{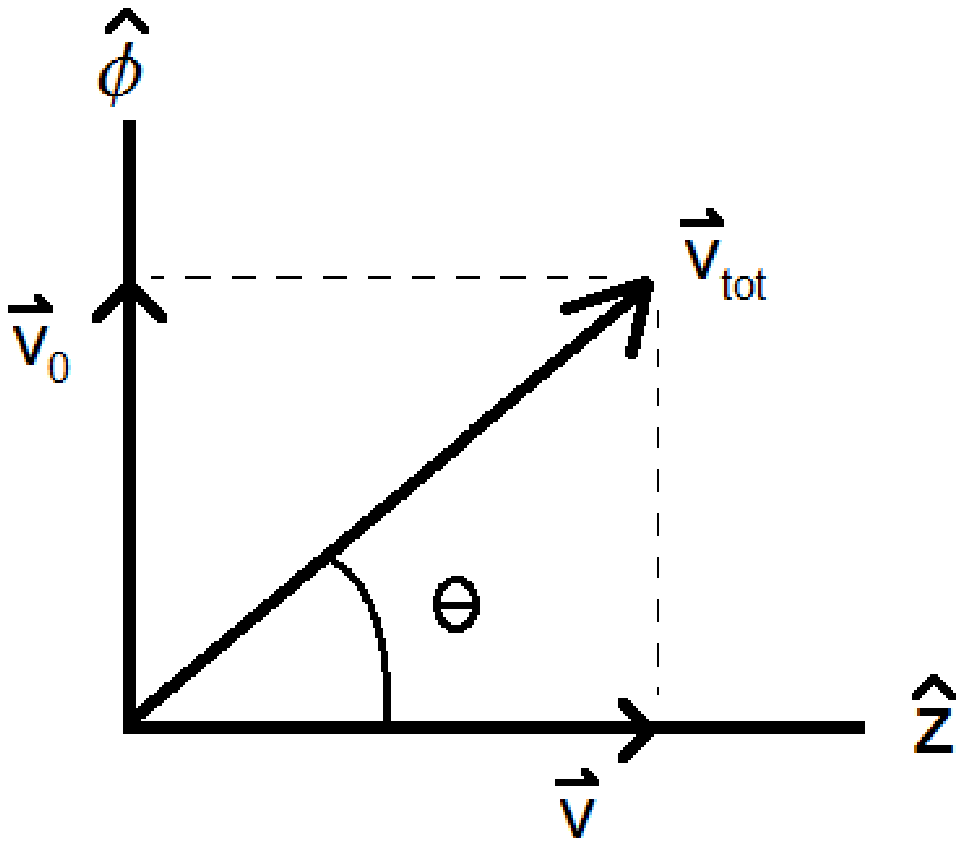}
\vspace{0.4cm}
\caption{(left) the decomposition of the total moment in its components $\vec{p}_0$ and $\vec{p}$; (right) the decomposition of the total velocity in its components $\vec{v}_0$ and $\vec{v}$. The axis $\hat{z}$ represents the Cartesian coordinate, whereas $\hat{\phi}$ represents a generic coordinate in the plane $x y$.}
\label{fig2}
\end{figure}

In Fig. \ref{fig2} we show the decomposition of $\vec{p}_{tot}$ in its components $\vec{p}_0$ and $\vec{p}$ (left figure) and the decomposition of $\vec{v}_{tot}$ in its components $\vec{v}_0$ and $\vec{v}$ (right figure). In the figures, $\vec{p}$ and $\vec{v}$ point out in the $\hat{z}$ Cartesian coordinate, whereas $\vec{p}_0$ and $\vec{v}_0$ point out in a generic coordinate designated by $\hat{\phi}$ which is contained in the plane $x y$.

\subsubsection{The modulus of the total velocity}

In Section \ref{rest} we proposed that, in the called ``rest frame'', the particle is doing a special microscopic motion with velocity $c$. This velocity is consistent with special relativity theory because we are considering that the ``mass'' is not an intrinsic property of the particle. The particle is intrinsically massless, and special relativity theory tells us that massless particles move with velocity $c$.

In this section we are considering the motion of the particle as seen from other inertial frames, in which the macroscopic velocity $\vec{v} \neq \vec{0}$. In order to be consistent with special relativity theory, we expect that the microscopic velocity $v_0$ is a decreasing function of $v$, so that the total velocity $v_{tot}$ is always equal $c$. If it is not the case, and $v_0$ is always equal $c$, when $v$ increases the total velocity $v_{tot}$ exceeds the light speed $c$, and the present theory becomes inconsistent with special relativity theory. In what follows we are going to derive the expression for $v_0$ and to verify if our expectation, i.e. $v_{tot} = c$, is satisfied.

The component $v_i$ of the velocity can be obtained by using the Hamilton equation:
\begin{equation}
\dot{q}_i = \frac{\partial H}{\partial p_i} \, ,
\end{equation}
where $\dot{q}_i$ is the time derivative of the coordinate $q_i$, $H$ is the Hamiltonian and $p_i$ is the canonical moment conjugated to $q_i$. For example, in the representation shown in Fig. \ref{fig2}, the macroscopic velocity of the particle is given by $\vec{v} = v \hat{z}$ and the moment conjugated to the coordinate $\hat{z}$ is the macroscopic moment $\vec{p}$. If we consider that the Hamiltonian of the particle is given by Eq. (\ref{ener13}) we obtain:
\begin{equation}
\dot{z} = \frac{\partial H}{\partial p} = \frac{\partial }{\partial p} \left( \sqrt{p_0^2 +  p^2} \, c \right) \,\, = \,\, \frac{p}{\sqrt{p_0^2 + p^2}} \, c \,\, = \,\, \frac{\left(\frac{mv}{\sqrt{1-v^2/c^2}}\right)}{\left(\frac{mc}{\sqrt{1-v^2/c^2}}\right)} \, c \,\, = \,\, v .
\label{vel1}
\end{equation}
This result is trivial, since in the representation shown in Fig. \ref{fig2} we have $v = \dot{z}$.

Now let us determine the microscopic velocity $v_0$. In the representation shown in Fig. \ref{fig2} we have: $\vec{v}_0 = v_0 \, \hat{\phi}$. Let us denote an infinitesimal displacement $\, d \, \vec{l} \,$ in the generic coordinate $\hat{\phi}$ as: $d \, \vec{l} = d \, l \,\, \hat{\phi}$. By using the Hamilton equation to determine $v_0$ we obtain:
\begin{equation}
v_0 = \frac{d l}{dt} = \frac{\partial H}{\partial p_0} = \frac{\partial }{\partial p_0} \left( \sqrt{p_0^2 +  p^2} \, c \right) = \frac{p_0}{\sqrt{p_0^2 + p^2}} \, c \, = \, \frac{mc}{\left(\frac{mc}{\sqrt{1-v^2/c^2}}\right)} \, c \, = \,  \sqrt{c^2 - v^2} \, .
\label{vel2}
\end{equation}
Therefore:
\begin{equation}
v_0 = \sqrt{c^2 - v^2} \, .
\label{vel3}
\end{equation}

Note that the velocity components $v$ and $v_0$ calculated above constitute the ``ordinary velocity'' and not the ``proper velocity'' of the particle \cite{relativ}. The proper velocity is that one in which the space intervals $dz$ and $dl$ travelled by the particle are measured by the observer, but the time interval used is the proper time $d\tau$ of the particle. The proper velocity componets are then given by $dz/d\tau$ and $dl/d\tau$. On the other hand, the ordinary velocity is that one in which the space intervals $dz$ and $dl$ as well as the time interval $dt$ are all of them measured by the observer. The proper velocity is used to define the 4-velocity in the Minkowski space \cite{relativ} and is more simple to Lorentz transform than the ordinary velocity, since the proper time $d\tau$ is Lorentz invariant and so only the space interval needs to be transformed. However, if you ask the observer to measure the particle velocity using his ruler and his clock the observer will obtain the ordinary velocity. 

Looking at Eq. (\ref{vel3}) we see that the microscopic velocity $v_0$ is a decreasing function of the macroscopic velocity $v$. Let us verify if our expectation concerning the total velocity $v_{tot}$ to be equal the light speed $c$ is satisfied. Since $\vec{v}_0$ is perpendicular to $\vec{v}$, the modulus of the total velocity $\vec{v}_{tot}$ is given by:
\begin{equation}
v_{tot} = \sqrt{v_0^2 + v^2} \, .
\label{vtotc}
\end{equation}
Substituting $v_0$ from Eq. (\ref{vel3}), we obtain: $v_{tot} = c$. Therefore, the total velocity of the ``massive'' fundamental particle is always equal the light speed $c$, as we expected. When the macroscopic velocity $v$ increases, the microscopic velocity $v_0$ decreases exactly the amount required to keep $v_{tot}$ always equal $c$. This result shows that the microscopic motion proposed in this paper is consistent with special relativity theory.

\subsection{How to do the Lorentz transformations in the present theory}
\label{Lorentr}

The formulas of the Lorentz transformations are well established, and the theory proposed in this paper does not imply any change in these formulas. However, an observation is in order. The present theory introduces an additional motion for massive fundamental particles. This additional microscopic motion, whose velocity is $\vec{v}_0$, is not known by special relativity theory. All the formulas of special relativity theory were constructed considering only the existence of the macroscopic velocity $\vec{v}$. Therefore, the correct way to do the usual Lorentz transformations is inserting the macroscopic velocity $\vec{v}$ in the formulas, and not the total velocity $\vec{v}_{tot}$ that was introduced by the present theory. For example, the dilation of a time interval is calculated through the formula \cite{relativ}: 
\begin{equation}
d\tau = \sqrt{1-v^2/c^2} \,\, dt \, .
\label{timedil}
\end{equation}
In this formula, we must insert the macroscopic velocity $\vec{v}$ instead of the total velocity $\vec{v}_{tot}$. It does not even make sense using $\vec{v}_{tot}$ in the Lorentz transformations, since $|\vec{v}_{tot}|=c$. In Eq. (\ref{timedil}), for example, if we insert $v_{tot}^2 = c^2$ we obtain $d\tau = 0$. Another example is the relativistic energy (\ref{ener3}). The velocity $v$ appearing in the denominator of Eq. (\ref{ener3}) is the macroscopic velocity, and not the total velocity.

Let us give an example to illustrate the use of the macroscopic velocity $\vec{v}$ to do the Lorentz transformations. In the previous subsection it was obtained the formula (\ref{vel3}), establishing that $v_0$ is a decreasing function of $v$. Instead of using the Hamilton equation to derive the formula (\ref{vel3}), we can use the Lorentz transformations to obtain it. According to special relativity theory, only the position vector parallel to the macroscopic velocity $\vec{v}$ is Lorentz contracted. The position vector perpendicular to $\vec{v}$ is not contracted. In the present theory it implies that, since the microscopic motion takes place in the plane perpendicular to $\vec{v}$, a space interval $d\vec{l}$ travelled by the particle along this plane is never Lorentz contracted. On the other hand, the proper time interval $d\tau$ of the particle, which is measured in the frame where $\vec{v}=\vec{0}$, is smaller than the time interval $dt$ measured by an observer situated in an inertial frame in which $\vec{v} \neq \vec{0}$, where $\vec{v}$ is the macroscopic velocity of the particle. So we have that $d\tau < dt$, and the relation between $d\tau$ and $dt$ is given by Eq. (\ref{timedil}). Therefore, in a frame where the macroscopic velocity $\vec{v} \neq \vec{0}$, the microscopic velocity $v_0$ is given by:
\begin{equation}
v_0 = \frac{dl}{dt} = \frac{dl}{d\tau} \sqrt{1-v^2/c^2} = c \, \sqrt{1-v^2/c^2} = \sqrt{c^2-v^2} \, ,
\end{equation}
where we used the fact that, in the ``rest frame'', the microscopic motion is done with velocity $c$, i.e., $dl/d\tau = c$.

This result shows us that the decrease of $v_0$ (when $v$ increases) can be interpreted as being a consequence of the Lorentz time dilation, which in turn is calculated inserting the macroscopic velocity $v$ in the formula (\ref{timedil}). The space interval $\, d \, \vec{l} \,$ travelled by the particle in the plane perpendicular to the macroscopic velocity $\vec{v}$ is the same in any inertial frame. However, the time spent by the particle to travel $\, d \, \vec{l} \,$ is different in different inertial frames. In the ``rest frame'', the time $d \tau$ spent by the particle to move $\, d \, \vec{l} \,$ is smaller than the time $d t$ measured by observers from other inertial frames seeing the particle moving the same space interval $\, d \, \vec{l} \,$.

\subsection{Geometrical interpretation for some results of special relativity theory}
\label{geomet}

The special relativity theory establishes that the velocity $v$ of a free massive particle can assume any value in the interval $0 \leq v < c$, but it can never be exactly equal $c$ because in the limit $v \rightarrow c$ the energy of the particle diverges. To verify this, it is enough to take a look at the formula (\ref{ener3}). When $v$ approaches $c$ the denominator of (\ref{ener3}) goes to zero, and consequently the energy $E$ goes to infinity.

The present theory provides a geometrical interpretation for this result. As discussed previously, the special relativity theory does not know the microscopic motion proposed in this paper. So, when special relativity theory tells us that the velocity of a massive particle is always smaller than the light speed it is referring to the macroscopic velocity $v$. It is the macroscopic velocity $v$ that is always smaller than $c$. To understand this result from a geometrical point of view, let us take a look at Fig. \ref{fig2}. Note that, even though the vectors $\vec{v}_{tot}$ and $\vec{p}_{tot}$ point out in the same direction, these two vectors have different properties. In the case of the velocity, the vector $\vec{v}_{tot}$ has constant modulus ($|\vec{v}_{tot}|=c$). Therefore, when the macroscopic velocity $v$ increases, the vector $\vec{v}_{tot}$ rotates clockwise. When it happens, the microscopic velocity $v_0$ decreases the amount required to keep the modulus of $\vec{v}_{tot}$ constant. In the case of the moment, it is the microscopic moment $\vec{p}_0$ that has constant modulus ($|\vec{p}_0|=mc$), and not the total moment $\vec{p}_{tot}$. In this case, a change in the macroscopic moment $p$ results in a change in the modulus of $\vec{p}_{tot}$ as well as in a rotation of this vector. However, in this process the microscopic moment $p_0=mc$ remains always constant.

Now let us analyse why the macroscopic velocity $v$ cannot be equal $c$. In Fig. \ref{fig2}, the right figure alone does not tell us that $v$ is prohibited to be equal $c$, but if we look at the left figure we see immediately the reason for this. The vector $\vec{p}_{tot}$ points out in the same direction as the vector $\vec{v}_{tot}$, so that the angle $\theta$ appearing in the left figure is the same angle $\theta$ appearing in the right figure. The point is that the vector $\vec{p}_0$ has a fixed modulus, $p_0 = mc$, what implies that the angle $\theta$ is always different of zero. A small angle $\theta$ can be reached when the macroscopic moment $p$ is very larger than the microscopic moment $p_0$, or equivalently when the total moment $p_{tot}=\sqrt{p_0^2+p^2}$ is very larger than $p_0$. However, the greater is the total moment $p_{tot}$ the greater is the energy of the particle $E=p_{tot}c$. In the limit $\theta \rightarrow 0$ (which corresponds to $v \rightarrow c$) the total moment $p_{tot}$ diverges, resulting in the divergence of the energy.

Therefore, from a geometrical point of view, what prohibits the macroscopic velocity $v$ to be equal $c$ is the fact that the modulus of the microscopic moment $\vec{p}_0$ does not depend on $v$, being a constant $p_0=mc$. It keeps the microscopic velocity $v_0$ always different of zero, even though $v_0$ can assume very small values when $v$ approaches $c$. The only way of having $v = c$ without the divergence of the energy is making $p_0 = 0$. In this case the angle $\theta$ becomes automatically zero, while the macroscopic moment $p$ remains finite. However, since $p_0 = mc$, when we make $p_0=0$ we are ``turning off'' the mass of the particle. In this case the particle ceases to do the microscopic motion ($v_0 = p_0 = 0$) and makes only the macroscopic motion with velocity $v = c$.

\section{About the microscopic motion}
\label{semi_classical}

\begin{figure}[htbp]
\includegraphics[scale=0.50]{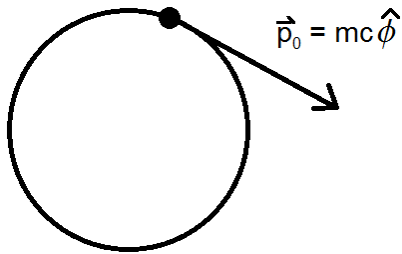}
\vspace{0.4cm}
\caption{The classical picture (classical in the sense of non-quantum mechanical) of a ``massive'' particle in orbital circular motion. The instantaneous linear moment of the particle, tangent to this motion, is $\vec{p}_0 = mc \, \hat{\phi}$, where $\hat{\phi}$ can be identified as the angular polar coordinate.}
\label{fig1}
\end{figure}

Until now we have proposed the existence of a microscopic motion without making any assumption about how is this motion. A very natural supposition is that it is a Microscopic Orbital Circular Motion (MOCM) [see Fig. \ref{fig1}]. 
In what follows we are going to analyse some characteristics of this hypothesis.

As discussed previously, the plane containing the microscopic motion is perpendicular to the macroscopic velocity $\vec{v}$. Consequently, if we suppose that the microscopic motion is the MOCM, the composition of the microscopic with the macroscopic motion will result in a helical motion. Here we make the observation that we are not saying that the particle does a helical trajectory because in quantum mechanics there is no trajectory. The correct is saying that there is a wave function describing the particle. This wave function is the composition of two waves, the plane wave associated with the macroscopic moment $\vec{p}$ and the wave associated with the MOCM.

\subsection{The classical radius of the MOCM}

The Heisenberg uncertainty principle implies that the fundamental particles, for example the electron, do not follow a classical trajectory. However, we can estimate the classical radius of the MOCM and consider this radius as being a characteristic length scale of the motion.

Let us take the angular moment $L$ associated with the MOCM (see Fig. \ref{fig1}):
\begin{equation}
L = r \, p_0 \, = \, r \, m c \, .
\end{equation}
From this relation, the classical radius `` $r$ '' of the MOCM is given by:
\begin{equation}
r = \frac{L}{m c} \, .
\label{raio_class}
\end{equation}

\subsubsection{The classical radius of the MOCM for the electron}

Let us suppose that the angular moment of the MOCM of the electron is $L = \hbar$, where $\hbar$ is the Plank constant. Substituting this value of $L$ in Eq. (\ref{raio_class}) we obtain the following result for the classical radius $r_e$ of the MOCM for the electron:
\begin{equation}
r_e = \frac{\hbar}{m c} \, \approx \, \frac{1.05 \times 10^{-34} \, J . s}{(9.1 \times 10^{-31} \, kg) \times (3 \times 10^8 \, m/s)} \, \approx \, 3.8 \times 10^{-13} \, m \, .
\label{re_el}
\end{equation}
Note that if we multiply $r_e$ by $2 \pi$ we obtain exactly the Compton wavelength of the electron.

\subsubsection{The frequency of the MOCM for the electron}

When the ``massive'' particle is in its ``rest frame'', it makes the MOCM with velocity $c$. So, if we know the classical radius $r$ of the MOCM we automatically know the frequency $\nu$ of this motion, i.e., we know how many times per second the particle completes one revolution, through the formula:
\begin{equation}
\nu = \frac{c}{2 \pi r} \, .
\label{freq}
\end{equation}
In the particular case of the electron, considering that the classical radius is given by Eq. (\ref{re_el}), we obtain:
\begin{equation}
\nu _e \approx \frac{3 \times 10^{8} \, m/s}{2 \times \pi \times (3.8 \times 10^{-13} \, m)} \approx 1.3 \times 10^{20} \, s^{-1}.
\end{equation}
As discussed in Subsection \ref{Lorentr}, when the macroscopic velocity $v$ increases the radius of the MOCM remains constant, since it is not affected by the Lorentz transformations. However, the microscopic velocity $v_0$ of the MOCM decreases when $v$ increases in consequence of the time interval dilation, being given by: $v_0 = \sqrt{c^2 - v^2}$. Since $v_0 = 2 \pi r \, \nu$, an observer situated in an inertial frame in which the macroscopic velocity $v \neq 0$ will see the particle doing the MOCM with a smaller frequency $\nu$.

\section{Summary}
\label{conc}

We proposed that there is a microscopic motion, done by all the massive fundamental particles, that is not explicit in the formulas of Special Relativity. The ``rest energy'' $E=mc^2$ is, ironically, an energy of motion. In sections \ref{rest} and \ref{outrest} we tried to answer all basic questions concerning this proposition in order to give a strong physical basis to it. We hope that this work can attract the interest of theoretical physicists who can give contributions to the development of this theory.


\begin{thebibliography}{99}



\bibitem{eletro}
  D.~Griffiths,
  ``Introduction to Electrodynamics'', $3rd$ ed.,
 Prentice Hall (1999); 
  J.~Jackson,
  ``Classical Electrodynamics'', $3rd$ ed.,
 John Wiley \& Sons, Inc. (1999).



\bibitem{relativ} 
W.~Rindler,
  ``Special Relativity'',
  Oliver and Boyd ltda., Edinburgh (1960). 




\end{thebibliography}
\end{document}